\documentclass[a4paper,fleqn,usenatbib]{mnras}
\usepackage[dvips]{graphicx}
\usepackage{colortbl,color}
\usepackage{amssymb,txfonts}
\usepackage[T1]{fontenc}
\usepackage{ae,aecompl}
\def\be{\begin{equation}}
\def\ee{\end{equation}}

\catcode`\@=11 
\def\@versim#1#2{\vcenter{\offinterlineskip
        \ialign{$\m@th#1\hfil##\hfil$\crcr#2\crcr\sim\crcr } }}

\begin{document}
\title[Hot accretion flow with radiative cooling]
{Hot accretion flow with radiative cooling: state transitions in
black hole X-ray binaries}
\author[M. C. Wu, et al.]
{Mao-Chun Wu$^{1}$\thanks{Email: maochun@ustc.edu.cn (MCW), fgxie@shao.ac.cn (FGX)};
Fu-Guo Xie$^{2}$; Ye-Fei Yuan$^{1}$;
Zhaoming Gan$^2$\\
$^{1}$Key Laboratory for Research in Galaxies and Cosmology,
Department of Astronomy,\\ University of Science and Technology of China, Hefei, Anhui, 230026, China;\\
$^{2}$Key Laboratory for Research in Galaxies and Cosmology, Shanghai Astronomical Observatory,\\
Chinese Academy of Sciences, 80 Nandan Road, Shanghai 200030, China}

\maketitle


\begin{abstract}
We investigate state transitions in black hole X-ray binaries
through different parameters by using two-dimensional axisymmetric
hydrodynamical simulation method. For radiative cooling in hot
accretion flow, we take into account the bremsstrahlung, synchrotron
and synchrotron-self Comptonization self-consistently in the
dynamics. Our main result is that the state transitions occur when
the accretion rate reaches a critical value $\dot M \sim 3\alpha\
\dot M_{\rm Edd}$, above which cold and dense clumpy/filamentary
structures are formed, embedded within the hot gas. We argued this
mode likely corresponds to the proposed two-phase accretion model,
which may be responsible for the intermediate state of black hole
X-ray binaries. When the accretion rate becomes sufficiently high,
the clumpy/filamentary structures gradually merge and settle down
onto the mid-plane. Eventually the accretion geometry transforms to
a disc-corona configuration. In summary our results are consistent
with the truncated accretion scenario for the state transition.

\end{abstract}


\section{INTRODUCTION}
The accretion of matter onto black holes, where a huge amount of
energy (both radiative and kinematic) is liberated out, is the key
process in black hole X-ray binaries (BHBs) and active galactic
nuclei (AGNs). Interestingly, it is argued that both BHBs and AGNs
can be divided into two different broad states/types. This is most
evident in BHBs, where two states with distinctive spectral and
timing properties are identified (see e.g. Zdziarski \& Gierlinski
2004; Remillard \& McClintock 2006; Belloni 2010 for recent reviews
on the state classifications and their observational properties). As
the source entering its outburst, generally it will first go through
a hard state, which is fainter but with a hard power-law spectrum.
The thermal component is highly suppressed in this state. As it
brightens, it will enter into a soft state, in which the spectrum is
characterized by a thermal component, supplemented with a weak
power-law tail. Between these two standard states, there also exist
a hybrid one, where the thermal and power-law components are
comparable. We call it intermediate state. In AGNs, although less
clear, more pieces of evidences are gathered recently to propose
that the bright AGNs are analogy to BHBs in their soft states, while
the low-luminosity AGNs (LLAGNs) are analogy to BHBs in their hard
states (e.g. Ho 2008; Antonucci 2012; Done 2014).

As widely accepted, the basic theoretical picture of these two
states is the truncated accretion -- jet model (Esin et al. 1997;
Yuan et al. 2005; see Yuan \& Narayan 2014 for the latest review).
In this model, the geometrically thin, optically thick, cold disc
(Shakura \& Sunyaev 1973; hereafter SSD) is believed to be truncated
at a certain radius $R_{\rm tr}$, inside which it is replaced by a
hot accretion flow such as the advection-dominated accretion flows
(ADAFs; Narayan \& Yi 1994, 1995; Abramowicz et al. 1995) or the
luminous hot accretion flow (LHAF; Yuan 2001, 2003; Yuan et al.
2007; Xie \& Yuan 2012). Additionally there is also a jet component,
which is believed to be connected to the hot accretion flow (Fender
et al. 2004; Wu et al. 2013). With this configuration,
observationally the system will be in the hard state. As the
accretion rate increases, $R_{\rm tr}$ becomes smaller. Eventually
when $R_{\rm tr}$ is smaller than the innermost stable circular
orbit (ISCO), i.e. the whole accretion is now a cold disc and the
system will be in the soft state. The readers are referred to Yuan
\& Narayan (2014) for the most recent review of the theory of hot
accretion flows, including its dynamics, radiation, and the
applications to observations.

Although non-radiative numerical simulations on hot accretion flows
have been investigated for years (hydrodynamic [HD]: e.g. Stone,
Pringle \& Begelman 1999; Igumenshchev \& Abramowicz 1999, 2000;
Yuan, Wu \& Bu 2012; magnetohydrodynamic [MHD]: e.g. Stone \&
Pringle 2001; Hawley, Balbus \& Stone 2001; Igumenshchev, Narayan \&
Abramowicz 2003; De Villiers et al. 2003; Machida et al. 2004;
Narayan et al. 2012; Yuan, Bu \& Wu 2012), only until very recently
the dynamical importance of radiative cooling on the hot accretion
flow has been examined through numerical simulations (e.g. Machida
et al. 2006; Fragile \& Meier 2009; Ohsuga et al. 2009;  Yuan \& Bu
2010; Ohsuga \& Mineshige 2011; Dibi et al. 2012; Li, Ostriker \&
Sunyaev 2013; S{\c a}dowski \& Narayan 2015). Moreover, the
theoretical interpretation of the state transition still lacks
direct support from numerical simulations. The first step, to our
knowledge, is taken by Das \& Sharma (2013), where they studied the
effects of radiative cooling on hot accretion flow. They found that,
there exist a certain critical density (depend on the viscous
parameter $\alpha$), above which the whole simulation domain will be
in a globally stable two-zone configuration, i.e. an outer cold disc
and an inner hot accretion flow. Such geometric configuration has
been expected for almost two decades (e.g. Esin et al. 1997).

Apart from its success, several shortages of their work could be
noted. Firstly, only bremsstrahlung emission is considered, while in
reality the innermost regions ($R< 20-30 \ R_{\rm s}$, where $R_{\rm
s}=GM_{\rm BH}/c^2$ is the Schwarzschild radius of the black hole;
cf. Fig.\ \ref{fig:runA} and Section\ 3.2.2) of hot accretion flow
are cooled by synchrotron and its Comptioization (Narayan \& Yi
1995; hereafter NY95). Secondly, they adopt a optically-thin
assumption and the radiative transfer is not taken into account,
which should not be the case when state transition occurs. Thirdly,
an one-temperature (ion temperature $T_{\rm i}$ equals electron
temperature $T_{\rm e}$ at every location) structure is assumed,
which is known to be incorrect (e.g. NY95; Moscibrodzka et al. 2009)
for most cases in hot accretion flows.

In this paper, we investigate the dynamical impacts of radiative
cooling in hot accretion flows by using two dimensional radiative HD
simulations, aiming at understanding the state transitions in BHBs
(and the evolution of AGN activity). The main improvement of our
work is that, besides the bremsstrahlung, both the synchrotron
radiation and its Comptioization are considered, following the
treatment of NY95\footnote{Note that, the NY95 approach is also
adopted in the general relativistic radiative MHD simulations by
Fragile \& Meier (2009). However, they assumed $T_{\rm e} = T_{\rm i}$, i.e. an
one-temperature accretion flow.}. Moreover, the impacts of gas
pressure to magnetic pressure ratio $\beta$, the ratio of electron
temperature to ion temperature $T_{\rm e}/T_{\rm i}$, on the
evolutionary stages of accretion flows, are also investigated.

Another motivation of this work is to verify, from state transition
simulations, the existence of LHAF (Yuan 2001, 2003; Yuan et al.
2007; Yuan \& Bu 2010; Xie \& Yuan 2012), a hot accretion flow
with accretion rate and radiative efficiency higher than those of
typical ADAFs. This new solution is promising for the understanding
of bright hard state (Yuan et al. 2007; Ma 2012). Besides, with
even higher accretion rate, analytical investigations of the
height-integrated one-dimensional solutions indicate that thermal
instability could be triggered. Consequently the accretion flow will
be possibly in two-phase (a type II LHAF branch, see Yuan 2003;
Xie \& Yuan 2012, 2016), i.e. numerous cold clumps will be
formed, embedded in the hot gas. This two-phase accretion flow may
relate to the intermediate state in BHBs (Yang et al. 2015).
Eventually as the accretion rate increases further, the cold clumps
may grow, merge and settle down to the mid-plane, resulting the
usual SSD. Together with the low-density hot medium above the cold
disc, the disc-corona configuration is recovered, and the accretion
system enters into a soft state.

The structure of this paper is as follows. In Section \ref{method},
we describe our numerical method and the treatment on various
radiative cooling mechanisms in hot accretion flow. The main results
are described in Section \ref{results}. Finally Section
\ref{summary} is devoted to a brief summary and discussions.

\section{METHOD} \label{method}

\subsection{Basic hydrodynamic equations}

In our numerical simulations, we use the ZEUS-2D code (Stone et al.
1992) to solve the basic hydrodynamic equations in spherical
coordinates ($R, \theta, \phi$):
\begin{equation}
\frac{\rm d\rho}{{\rm d} t}+\rho\nabla\cdot
\mathbf{v}=0,\label{eq:cont}
\end{equation}
\begin{equation}
\rho\frac{\rm d\mathbf{v}}{{\rm d} t}=-\nabla P_{\rm tot}-\rho\nabla
\phi+\nabla\cdot\mathbf{T}, \label{eq:mon}
\end{equation}
\begin{equation}
\rho\frac{\rm d (e/\rho)}{{\rm d}t}=-P_{\rm
gas}\nabla\cdot\mathbf{v}+\mathbf{T}^2/\mu-Q^{-}_{\rm rad}.
\label{eq:energy}
\end{equation}
Here $\rho$ is the mass density, ${\bf v}$ is the gas velocity,
$P_{\rm tot}\equiv P_{\rm gas}+P_{\rm mag}$ is the total
(gas+magnetic) pressure, $\bf T$ is the anomalous stress tensor. A
pseudo-Newtonian gravitational potential (Paczy{\'n}ski \& Wiita
1980) is adopted, $\phi=-GM_{\rm BH}/(R-R_{\rm s})$. Besides,
for an ideal gas, the internal energy $e$ can be expressed as, $e =
P_{\rm gas}/(\gamma - 1)$, where $\gamma = 5/3$.

The stress tensor $\bf T$ is now widely believed to be associated
with MHD turbulence driven by the magneto-rotational instability
(MRI, see Balbus \& Hawley 1998 for a review). However, since our
simulation is hydrodynamic, we follow the conventional strategy, to
use the $\alpha$-viscosity description (Shakura \& Sunyaev 1973),
Following Stone, Pringle \& Begelman (1999), we include the non-zero
azimuthal components of $\bf T$,
\begin{equation}
  T_{\rm {r\phi}} = \mu R \frac{\rm \partial}{{\rm \partial} R}
    \left( \frac{v_{\phi}}{R} \right),
\end{equation}
\begin{equation}
  T_{\rm {\theta\phi}} = \frac{\mu \sin \theta}{R} \frac{\rm \partial}{\rm \partial
  \theta} \left( \frac{v_{\phi}}{\sin \theta} \right).
\end{equation}
Here, the viscous coefficient $\mu=\nu\rho$ and the kinematic
coefficient $\nu \propto r^{1/2}$, which is similar to the standard
$\alpha$-description (Shuakura \& Sunyaev 1973).

Although global large-scale magnetic field may exist in the hot
accretion flow, we here neglect it since our simulations are
hydrodynamic (see Section 5 for discussions on this point.). The
local and randomly-oriented (tangled) magnetic field is determined
through the plasma $\beta$ parameter  $\beta=P_{\rm gas}/P_{\rm
mag}$. Consequently the total pressure can be expressed as $P_{\rm
tot}=P_{\rm gas}+P_{\rm mag} = (1+\beta)/\beta\ P_{\rm gas}$. MHD
numerical simulations of hot accretion flow indicate that, the value
of $\beta$ is not a constant, but varies with time and location
($r,\theta$) within the simulation domain. Moreover, the average
value of $\beta$ likely depends on the magnitude of the net initial
magnetic field (e.g. Stone \& Pringle 2001; Hawley \& Krolik 2001;
Hawley et al. 2001; Beckwith et al. 2008. See Yuan \& Narayan 2014
for review.). For simplicity we assume $\beta$ is a constant.

\begin{figure}
\begin{center}
\includegraphics[width=0.45\textwidth]{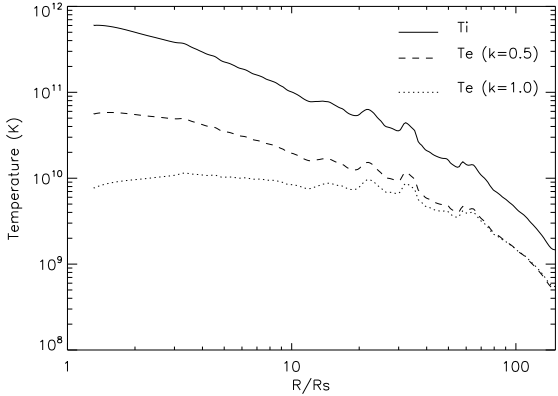}
\end{center}
\caption{Ion temperature $T_{\rm i}$ and electron temperature
$T_{\rm e}$ as a function of radius for cases with $k=0.5$ and
$k=1$. The ion temperature is taken in fiducial {\bf Run A} (see
Table 1) at a quasi-steady state before considering radiative
cooling.}\label{fig:init}
\end{figure}

The hot accretion flow is two-temperature (NY95), with electron
temperature is much lower than ion in the innermost regions of the
accretion flow, due to the fact that both the strong radiative
cooling of electrons and different adiabatic index between electrons
and ions (e.g. Narayan, Mahadaven \& Quataert 1998; Yuan \& Narayan
2014). We use the following formulae to mimic the results from
height-integrated numerical calculations (NY95; Xie et al. 2010),

\begin{equation}
T_{\rm e} = {T_{\rm i}\over(R/R_0)^{-k} +2.0},
\end{equation}
where $k>0$ is a tuning parameter and $R_0$ is the density maximum
location of the initial torus (see Section\ \ref{sec_ini} for the
torus setup). In other words, we assume $T_{\rm e} \approx 0.3
T_{\rm i}$ at large radii ($r>100\ R_{\rm s}$), and $T_{\rm e}$ can
be highly suppressed (compared to $T_{\rm i}$) within $\sim 20 \
R_s$, where most of the radiation are emitted. Fig. \ref{fig:init}
shows $T_i $ and $T_e$ as a function of radius for $k= 1$ and $k=0.5
$, the ion temperature plotted here is when the accretion flow gets
to a quasi-steady state without adding radiative cooling. For a
direct comparison, the radiative cooling of the bremsstrahlung
emission at larger radii will be about a factor of $\sqrt{T_{\rm
i}/T_{\rm e}}\approx \sqrt{3}$ lower, compared to those
one-temperature simulations. On the other hand, for the dense
clumps/filaments in our simulations, we underestimate the electron
temperature because the clumps will be nearly one-temperature
because of the strong energy exchange/coupling between electrons and
ions.

\subsection{Radiative cooling rate $Q^{-}_{\rm rad}$}

We now provide the numerical formulae to derive the radiative
cooling rate $Q^{-}_{\rm rad}$. Technically there are several
treatments on the radiation, e.g. some researchers employ the
flux-limited diffusion approximation to solve the radiation energy
equation (Ohsuga et al. 2009). Here we take a different and also
simplified approach to calculate all the radiative cooling terms. We
note that the radiative processes in hot accretion flow include the
bremsstrahlung, synchrotron, and the Comptonization (the seed
photons are mainly the synchrotron photons, whose energy is much
lower compared to that of bremsstrahlung photons.). The radiative
cooling rate can then be expressed as,
\begin{equation}
Q^-_{\rm rad} = Q^-_{\rm brem} + Q^-_{\rm syn}+ Q^-_{\rm syn, C}.
\end{equation}
Here, $Q^-_{\rm brem}$ is the bremsstrahlung emission including both
the electron-ion and electron-electron collisions (Eq. 3.4 in NY95).

Optically thin synchrotron radiation is emitted mainly at the local
self-absorption peak frequency $\nu_c$, below which the synchrotron
emission is self-absorbed. Once the $\nu_c$ is determined at each
location, we adopt a simplified formula (Eq. 3.18 in NY95) to
estimate the synchrotron emission,
\begin{equation}
Q^-_{\rm syn}\approx{2\pi\over 3 c^2}kT_{\rm e}(R){\nu_c^3(R)\over R}.
\end{equation}
The above formula ensures that the integral of $Q^-_{\rm syn}$ over
the entire accretion flow roughly equals to the total cooling
radiation that reaches infinity (NY95). For the estimation of
$\nu_c$, we also follow the approach of NY95, with a slight
modification. At radius $R$, the emission volume is revised to
$2\times H \times \Delta S$, where $H$ is the scale height and
$\Delta S$ is the surface area, instead of $4/3\ \pi R^3$. We then
equate the synchrotron emission to the Rayleigh-Jeans blackbody
emission from the surface of that volume. The net effect of this
modification is to replace the $R$ of eq. 3.14 in NY95 with $3 H$,
i.e. the surface density plays its role in determining $\nu_c$.

One key process is the Compton scattering within the hot accretion
flow. In this work we use the Compton enhancement factor $\eta$
(Dermer, Liang \& Canfield 1991) which is determined by the electron
energy, the seed photons, the geometry of the Compton region and
also the effective optical depth $\tau_{\rm eff}$. In the
calculation of $\tau_{\rm eff}$, we adopt coefficients that
correspond to a disc configuration and seed photon energy at 1 eV
(Dermer et al. 1991; Eq. 3.19 in NY95). Moreover, since we do not do
any integrations in the calculation of either synchrotron or
Comptonization, we take $\nu_c$ as the mean seed photon energy. The
radiative cooling rate due to the Comptonized synchrotron emission
is,
\begin{equation}
Q^-_{\rm syn, C} = (\eta(T_{\rm e}, \nu_c, \tau_{\rm eff})-1)\ Q^-_{\rm syn}.
\end{equation}
Two notes should be emphasized. First, all these formulae are based
on the local Comptonization approximation which is crude and
additional corrections should be made (e.g. Yuan, Xie \& Ostriker
2009; Xie et al. 2010; Nied{\'z}wiecki et al. 2012). Second, the
radiative transfer of the optically thin medium has been
approximately taken into account under this approach, although we
can not quantitatively estimate how goodness it is. We devote it to
future work.

Since we investigate accretion flow with moderately high accretion
rate, so cold clumps with large optical depth is expected. In order
to take this effect into account, we define $\tau_{\rm es}$ the
electron scattering optical depth in $\theta$ direction, from disc
midplane $\theta=\pi/2$ to the local location $\theta$. And also
$\tau_{\rm abs} $ is the absorption optical depth. If the value of
$\tau_{\rm abs}$ exceeds $\sim 1/3$ (arbitrarily set), we will
correct the high-optical-depth issue, the following formulae is
adopted (Hubeny 1990; see also NY95),
\begin{equation}
Q^-_{\rm rad} = {4\sigma T_{\rm e}^4\over H}\ {1\over\left(3\tau/2+\sqrt{3}+1/\tau_{\rm abs}\right)}.
\end{equation}
 Here $\tau = \tau_{\rm abs} + \tau_{\rm es}$ is the total optical depth.
 $\tau_{\rm abs}$ can be approximated as (cf. Eq. 3.33 in NY95),
\begin{equation}
\tau_{\rm abs} = {H\over 4\sigma T_{\rm e}^4}\ \left(Q^-_{\rm brem} + Q^-_{\rm syn}+ Q^-_{\rm syn, C}\right).
\end{equation}

\subsection{Initial conditions}\label{sec_ini}

The initial setup is an equilibrium torus around a stellar mass
black hole, with mass $M_{\rm BH}=10\ M_{\odot}$. The torus
structure is described as (Papaloizou \& Pringle 1984),
\begin{equation}
 \frac{P_{\rm gas}}{\rho} = \frac{(\gamma-1)GM_{\rm BH}}{\gamma R_{0}} \left[ \frac{R_{0}}{R} - \frac{1}{2}
\left( \frac{R_{0}}{R \sin \theta} \right)^{2} - \frac{1}{2d}
\right] .
\end{equation}
Here $R_{0}= 100 R_{\rm s}$ is the radius of the torus center
(density maximum $\rho_{\rm max}$), $d$ is the distortion of the
torus. Additionally the torus is embedded in a low density medium.
The density of these ambient gas density is $\rho_{\rm amb} = 1
\times 10^{-4}\ \rho_{\rm max}$. The specific angular momentum of
the tours equals to the Keplerian angular momentum at $R_0$. We
include the randomly-oriented magnetic fields in the simulations
through plasma parameter $\beta$.

Different parameters are used to depict the accretion flow. The
viscosity parameter $\alpha$ is fixed to 0.01, which is a typical
value of current numerical simulations (e.g. Hawley et al. 2011;
Mckinney et al. 2012); The ratio of gas pressure to magnetic
pressure $\beta$; The maximum density $\rho_{\rm max} $ in the torus
center which determines the quantity $\dot{M}$ in units of Eddington
accretion rate ($\dot{M}_{\rm Edd} \equiv 10\ L_{\rm Edd}/c^2$,
where $L_{\rm Edd}$ is the Eddington luminosity); Parameter $k$
denotes the ratio of electron temperature to ion temperature. The
parameters are summarized in table 1, columns (2) to (6) give the
maximum density in the center of torus (in units of $\rho_0$,
$\rho_0\equiv 0.77 \times 10^{-6} \rm g \cdot cm^{-3}$), plasma
parameter $\beta$, $k$, the total run time of the simulation (in
units of orbital time at radius of $R_0= 100 R_{\rm s}$), and mass
accretion rate $\dot M$ (in units of $\dot{M}_{\rm Edd}$).

\begin{table}
\footnotesize
\begin{center}
\caption{Model Parameters and Simulation Properties}\label{tab}
\begin{tabular}{ccccccc}
\hline
Run   & $ \rho_{\rm max} / \rho_0$ & $ \beta $ & $ k $ & $t_{\rm stop}\ ^{(a)}$& $\dot{M}/\dot{M}_{\rm Edd}\ ^{(b)}$ \\
\hline
A    & 1.0   & $ 10    $  & 1.0     & 9.0  & 0.06  \\
A1   & 0.5   & $ 10    $  & 1.0     & 8.0  & 0.03  \\
A2   & 0.1   & $ 10    $  & 1.0     & 5.0   & 0.006 \\
B    & 1.0  & $ 100     $  & 1.0     & 5.0   & 0.06   \\
B1   & 1.0  & $ 1   $  & 1.0   & 5.0   & 0.06  \\
E    & 0.5  & $ 10     $  & 0.5   & 5.0   & 0.03 \\
E1   & 0.5   & $ 10    $  & ...$^{(c)}$   &5.0   & 0.03  \\

\hline
\end{tabular}\\
\end{center}

(a) The whole physical simulation time $t_{\rm stop}$ in unit of the orbital period at $R_0 = 100\ R_{\rm s}$.\\
(b) The net accretion rate at location $2\ R_s$ at time $1.4 $
orbits,
when the whole accretion flow is in a quasi-steady state.\\
(c) In this case, we set $T_{\rm e} = T_{\rm i} $.
\end{table}

\subsection{Numerical methods}

The public ZEUS-2D code (Stone \& Norman 1992) are used in this work
to solve the basic hydrodynamic equations. The shear stress and the
radiative cooling are considered. The radiative cooling term is
treated explicitly. The Courant--Friedrichs--Lewy condition for the
radiative cooling is not taken into account. Instead, a sub-cycle
technique is adopted for regions with high density (in the cold
clumps), i.e. whenever the radiative cooling time-step is smaller
than the time-step used for the hydrodynamic equations, we sub-cycle
that. The radiative terms are repeatedly at the smaller time-step
until one hydrodynamical time-step has elapsed for the purpose of
saving computational time.

In our simulations, the inner and outer boundary (in radial
direction) of the computational domain are set as, $R_{\rm in}=1.3\
R_{\rm s}$ and $R_{\rm out}= 400\ R_{\rm s}$. We use outflow
boundary conditions at these two boundaries so that mass is allowed
to flow out of the computational domain but not into it. In the
angular direction, the axis symmetric boundary condition is taken.

We followed the treatment of Stone et al. (1999) for the grid setup,
i.e. logarithm in both the radial and the angular directions. More
specifically, we choose in radial direction $(\bigtriangleup
r)_{i+1} / (\bigtriangleup r)_{i} = \sqrt[N_{r}-1]{10}$ with $N_{r}$
grid points per decade in radius. To resolve the cold clumps (and
also the possible think disk) near the equator, we adopt non-uniform
angular zones with $(\bigtriangleup \theta)_{j} / (\bigtriangleup
\theta)_{j+1} = \sqrt[N_{\theta}-1]{4}$ for $0 \leq \theta \leq
\pi/2$, and $(\bigtriangleup \theta)_{j+1} / (\bigtriangleup
\theta)_{j} = \sqrt[N_{\theta}-1]{4}$ for $\pi/2 \leq \theta \leq
\pi$. This gives a refinement by a factor of 4 in the angular grid
zones between the poles and equator. The standard resolution is
$N_{\rm r} =128$ and $N_{\rm \theta} =80$, giving a grid with a
total size of $334\times160$ zones. \textbf{The standard resolution
gives the minimum grid of $\Delta r = 0.015 R_{\rm s}$ and $\Delta
\theta = 0.5\degr$}. We have also computed a high resolution model
with $N_{r} \times N_{\theta}=128 \times 160$ (total $334\times
320$). We caution that although the resolution is high enough to
resolve the cold clumps formed near the equatorial plane, it is
still insufficient to resolve clumps possibly formed at high
altitude. Simulation with higher resolution is computationally
expensive, which is beyond the scope of current work.

\begin{figure*}
\begin{center}
\includegraphics[width=0.90\textwidth]{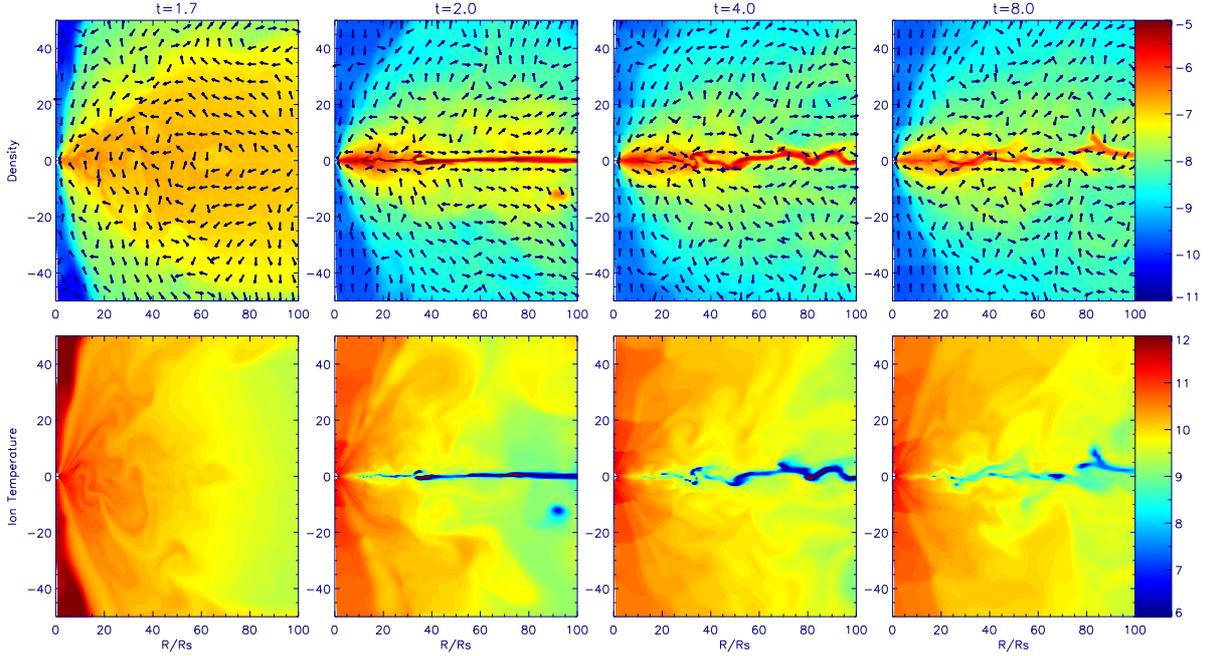}

\end{center}
\caption{The logarithmic density (over-plotted with poloidal
directions of the velocity as arrows) and temperature, for {\bf Run
A}. Panels from left to right show time at $t = 1.4, 2.0, 4.0, 8.0
$, respectively.}\label{fig:runA}
\end{figure*}
Technically, we first run the simulation (of the equilibrium torus)
without radiative cooling. After the accretion flow enters a quasi-steady state,
we then start to include the radiative cooling.

\section{RESULTS}\label{results}

As summarized in Table 1, we have run seven HD simulations with
different model parameters. We take {\bf Run A} ($\alpha=0.01$,
$\beta=10$, $k= 1.0$) as the fiducial run. The accretion time is in
the units of the orbital period at radius $R_0 = 100\ R_{\rm s}$ in
all models. We will first present results of this fiducial model to
provide a general picture of our simulations, and then will go
further to discuss the influences of individual model parameters.
Note that the dynamical impact of the viscosity parameter $\alpha$
on hot accretion flows is well-known and easy to understand (see
e.g. Stone, Pringle \& Begelman 1999; Igumenshchev \& Abramowicz
1999, 2000 for cases without radiation; Yuan \& Bu 2010; Das \&
Sharma 2013 for cases with radiation), and we omit to discuss it
here.

\subsection{Fiducial {\bf Run A}}
When the accretion flow reaches a quasi-steady state ($t =1.4 $
orbits), a representative of hard state in BHBs, we start to include
the radiative cooling. Very quickly the state transition is
triggered. If initially we setup with a very low density torus and
the radiative cooling is weak (dynamically un-important), then such
state transition will not happen at all. At a later time, $t = 2.0$,
the whole accretion flow is now dominated by a cold, geometrically
thin accretion disc and surrounded by a hot tenuous media, similar
to the disc-corona configuration. This is likely a representative of
the soft state. For completeness, we also present the results at $t
= 4.0$ and $t = 8.0 $, where the mass accretion rate is reduced
gradually, because there is no continuous mass supply in our
simulations.

\begin{figure}
\begin{center}
\includegraphics[width=0.45\textwidth]{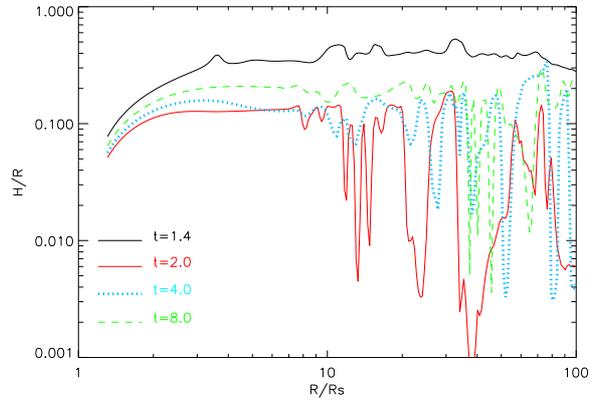}
\end{center}
\caption{The aspect ratio H/R ($H \approx c_{\rm s} /\Omega_{\rm
K}$) at the equatorial plane versus radius for {\bf Run A}. The
black solid, red solid, blue dotted and green dashed curves are
respectively, correspond to the time $t= $ 1.4, 2.0, 4.0,
8.0.}\label{fig:hr}
\end{figure}

\begin{figure}
\begin{center}
\includegraphics[width=0.45\textwidth]{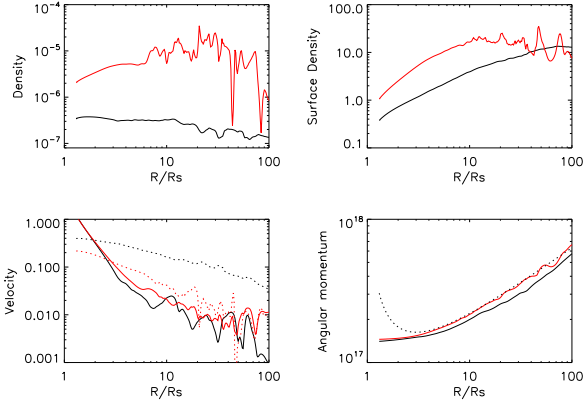}
\end{center}
\caption{Radial scaling of time-averaged physical quantities for
{\bf Run A}. From left to right, upper to lower are the density,
surface density, velocity and angular momentum, respectively.In each
panel, colors denote different evolution time, the black color lines
denotes $t$ = 1.4 and red color line denotes $t$ = 2.0. The density,
surface density and angular momentum are in units of cgs. Only the
velocities are in units of light speed.  In the velocity panel, the
solid lines denote radial velocity and dotted lines denote sound
speed. In the angular momentum panel, dotted line represents the
Keperlian angular momentum.}\label{fig:prof}
\end{figure}

\begin{figure*}
\begin{center}
\includegraphics[width=0.85\textwidth]{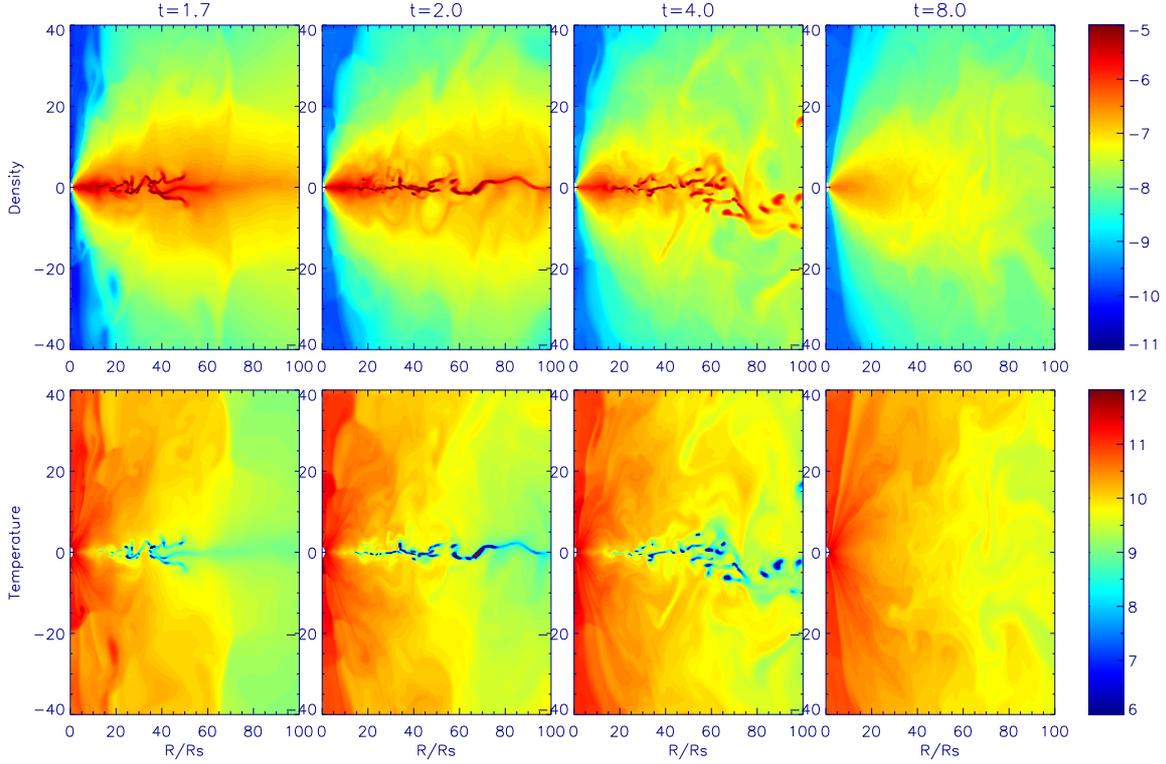}
\caption{The density (upper panels) and ion temperature (lower
panels) in logarithmic scale for {\bf Run A1} (cf. Table 1). The
panels show at different time, from left to right, $t=$ 1.7, 2.0,
4.0, 8.0, respectively.}\label{fig:A1run}
\end{center}
\end{figure*}

To better understand the detailed structure of the accretion flow,
Fig. \ \ref{fig:runA} shows the density (upper panels) and ion
temperature (bottom panels) for the four representative times as
mentioned before and the density panels over-plotted with poloidal
velocity field. At time $t=1.4$, the accretion flow is globally
geometrically thick with high temperature. Moreover, it is turbulent
inflow near the equatorial plane, and outflow above certain altitude
(see Yuan, Wu \& Bu 2012 and references therein). At time $t = 2.0$,
the cold, dense, geometrically thin accretion flow extends to radii
$R>100\ R_s$. After this stage, the dynamical structures of thin
disc becomes turbulent as shown in the right two panels of Fig.
\ref{fig:runA}. This is because there is no continuous mass supply
in our simulations. Consequently as the system evolves, the gas
density reduces, and the radiative cooling becomes less important.
In other words, if we run the simulation for sufficiently long time,
the accretion flow will eventually return back to a hot accretion
flow, i.e. a transition from soft state to hard state. Note that, a
soft-to-hard state transition is realized in {\bf Run A1}, where the
initial density and consequently the accretion rate are lower (cf.
Fig. \ref{fig:A1run}).

Fig.\ \ref{fig:hr} shows the ratio of disc scale-height to radius
$H/R$, as a function of radius. The disc scale-height is defined as,
$H \approx c_s/\Omega_{\rm K}$, where $c_s$ is the sound speed and
$\Omega_{\rm K}$ is the Keplerian angular velocity. For simplicity,
we only calculate the aspect ratio at the mid-plane. The black
solid, red solid, blue dotted and green dashed curves show $H/R$ at
time $t=$ 1.4, 2.0, 4.0 and 8.0, respectively. For technical reasons
we take the accretion flow with $H/R > 0.1$ as hot accretion flow
and $H/R < 0.01$ as cold disc. From this definition, the accretion
flow is a hot, geometrically thick disc at time $t=1.4$, where the
truncated radius $R_{\rm tr} > 100\ R_{\rm s}$, this stage is
corresponded to the hard state.  At time $t = 2.0 $ when radiative
cooling is taking into account, numerous cold components are formed
(see also Fig.\ \ref{fig:runA}), almost represents a cold disc which
is truncated at radius $\approx 10\ R_{\rm s}$. Inside the truncated
radius the accretion flow remains hot. As accretion rate decreases,
the truncated radius become larger gradually, $R_{\rm tr}\approx 25
\ R_{\rm s}$ at time $t=4.0$ and $R_{\rm tr}\approx 35 \ R_{\rm s}$
at time $t=8.0$. This results is consistent with previous works
(e.g. Das \& Sharma 2013).

 Fig.\ \ref{fig:prof} shows the time
averaged radial structures of the accretion flow at the equatorial
plane, time is averaged over a duration of 0.1 orbits. In each
panel, the black and red lines correspond to time  $t= 1.4$ (the
typical non-radiative hot accretion flow, likely the hard state) and
$t = 2.0 $ (the radiative accretion flow, likely the soft state),
respectively. It is clear that each variable at hard state can be
described by a simple radial power-law distribution, the density
scales as $\rho \propto r^{-0.3}$, the surface density scales as
$\Sigma \propto r^{-3/2}$, the radial velocity scales as $v_{r}
\propto r^{-3/2}$ and the angular momentum scales as $l \propto
r^{0.5}$.

\begin{figure}
\begin{center}
\includegraphics[width=.45\textwidth]{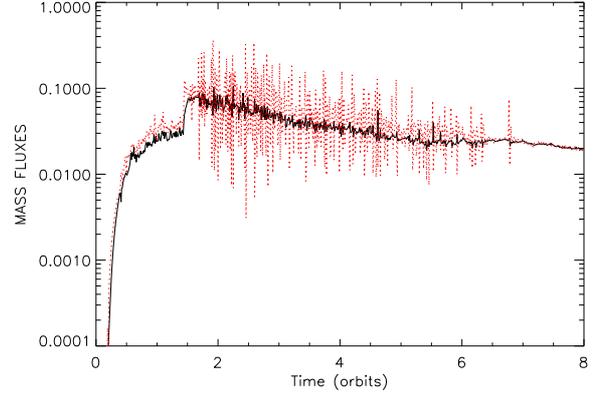}
\caption{Mass accretion rate $\dot M$ (in units of Eddington
accretion rate) at $R=2\ R_{\rm s}$ (black solid line) and $R=10\
R_{\rm s}$ (red dotted line), as a function of time (in units of the
orbital time at $R_0$) for {\bf Run A1}.}\label{fig:mdota1}
\end{center}
\end{figure}

\begin{figure}
\begin{center}
\includegraphics[width=0.45\textwidth]{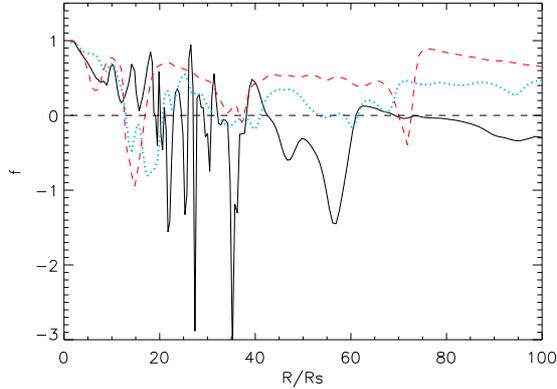}
\caption{The snapshot ($t=1.7$) of advection factor $f$ for {\bf Run
A1} at different angular locations, $\theta = 90\degr$ (black solid
line; equatorial plane), $\theta = 70\degr$ (blue dotted line) and
$\theta = 50\degr$ (red dashed line)
respectively.}\label{fig:energya1}
\end{center}
\end{figure}
\begin{figure*}
\begin{center}
\includegraphics[width=0.45\textwidth]{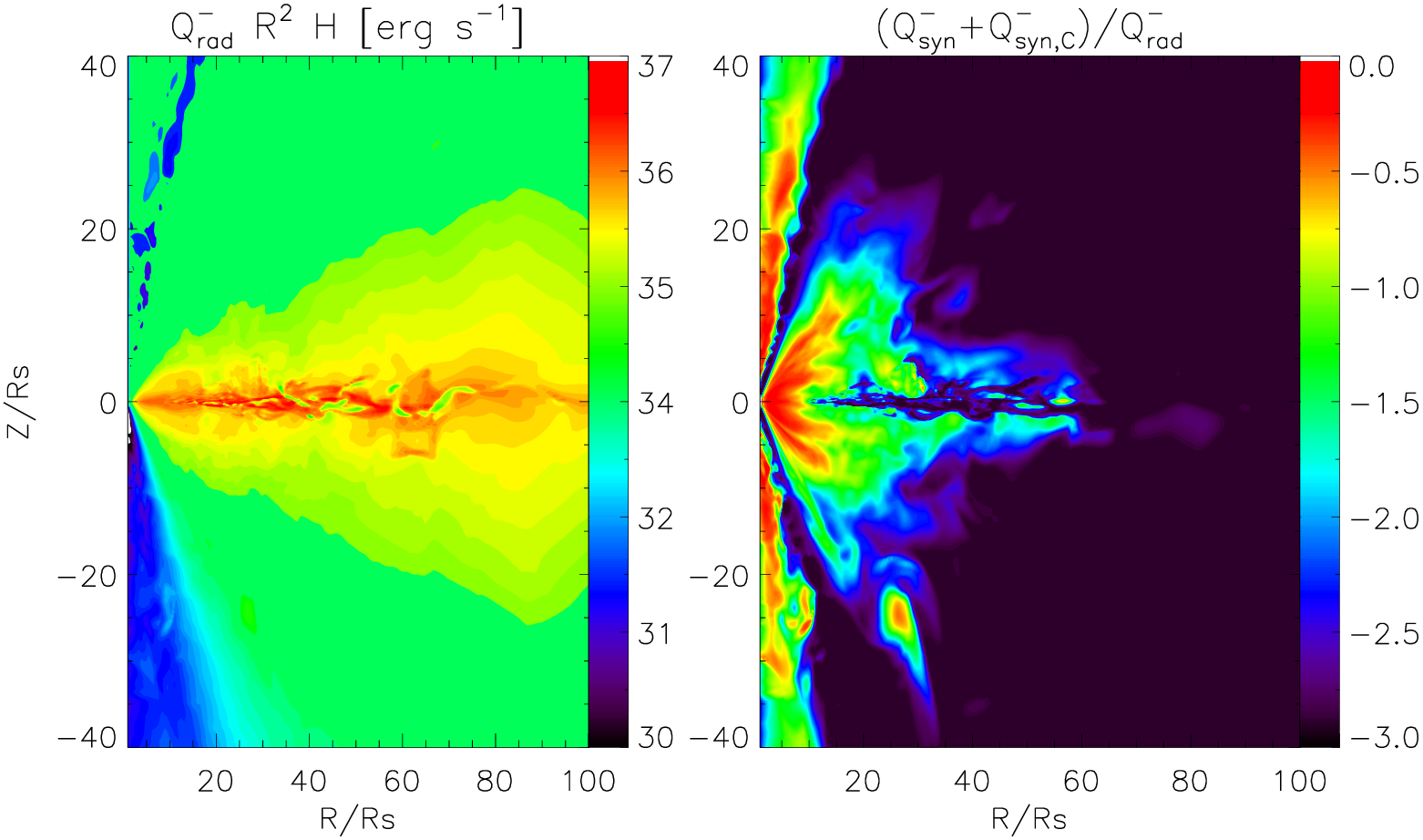}\hspace*{1.0cm}
\includegraphics[width=0.45\textwidth]{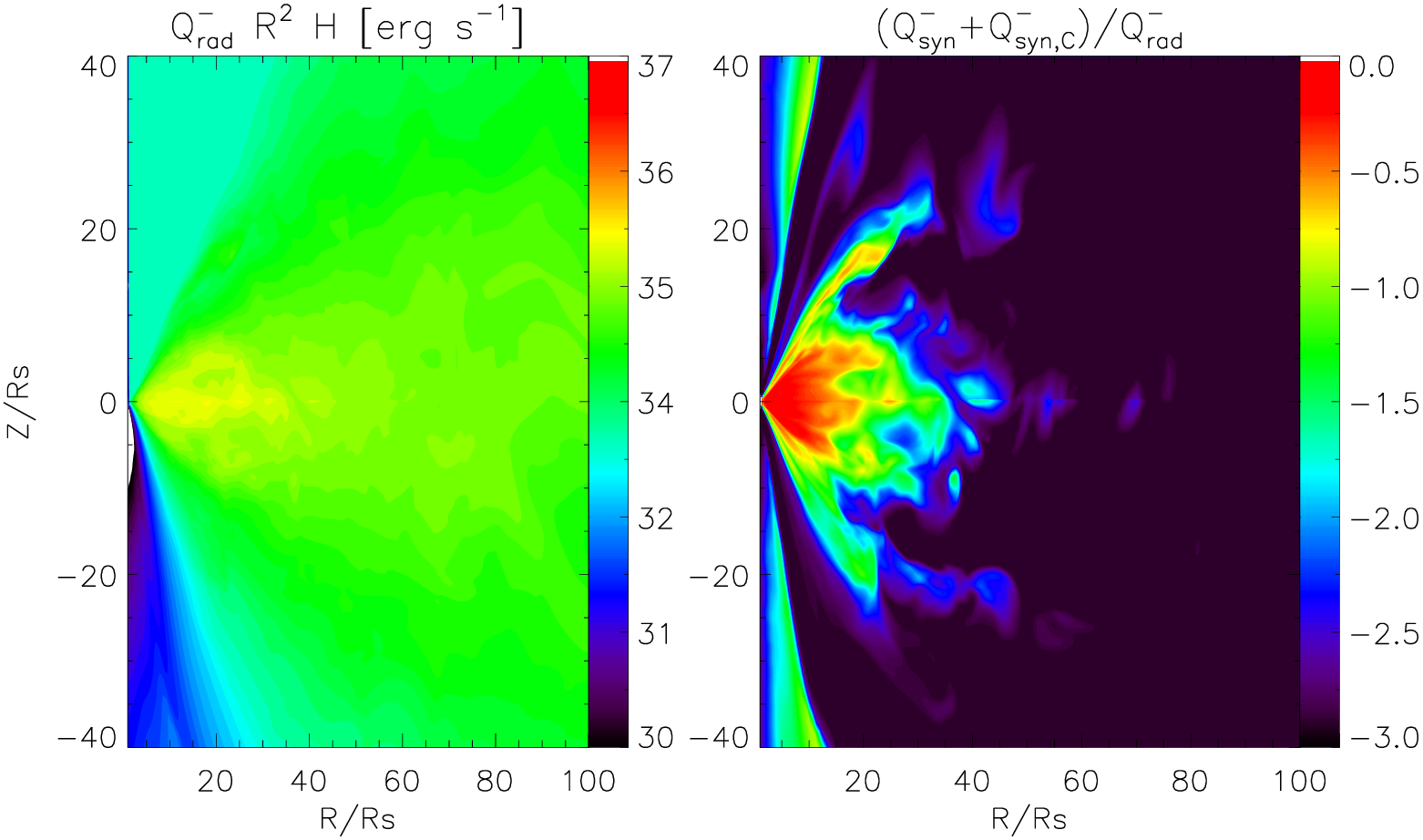}
\end{center}
\caption{The logarithmic radiative cooling term for {\bf Run A1}
over the period $1.8 - 2.0$ (left two panels) and $7.8 - 8.0$ (right
two panels). The first/third panel shows the total radiative cooling
rate multiplied by $R^2\ H$ and the second/fourth panel shows ratio
between Comptonized synchrotron radiative cooling rate and radiative
total cooling rate.}\label{fig:radA1}
\end{figure*}
After considering of the radiative cooling, the accretion flow
change significantly in comparison with non-radiative model. The
density increases by nearly two orders of magnitude, while the
surface density increases by only a factor of $\sim 10$ due to the
reduction in the disc thickness. Besides, it should be point out
that the cold disc in our simulation is highly different from the
standard SSD model, although its rotational velocity is Keplerian
(bottom right panel in Fig.\ \ref{fig:prof}), the profiles in
density, surface density, temperature as well as velocities deviate
from predictions in SSD (Shakura \& Sunyaev 1973; Frank, King \&
Raine 2003). Most notably, the radial velocity is very high in cold
accretion flow. There are several reasons for these results.
Firstly, the opacity we adopt for the cold disc is not the Kramers'
law. Secondly, the treatment for the optically thick disc is still
very crude. We expect it will eventually become the SSD, if several
technical problems are solved, i.e. the radiative cooling will be
well-treated, and more importantly, the resolution (especially in
the angular direction) will be highly improved. Global simulations
of SSD is beyond the scope of this work.

\subsection{The dependence on $\dot M$ }
The first and most apparent parameter is the accretion rate, or
density of the initial torus, which will control the whole mass
supply of the system. To investigate its effect, we run another two
simulations with different accretion rate including {\bf Run A1}
with $\dot M = 0.03\ \dot M_{\rm Edd}$ and {\bf Run A2} with  $\dot
M = 0.006\ \dot M_{\rm Edd}$. We use a slightly high numerical
resolution (total: $334\times 320$) in {\bf Run A1} to discern
filament structures.  Our simulation results indicate that the
critical mass accretion rate is about $\sim  0.03\ \dot M_{\rm
Edd}$, below this critical accretion rate the hard-to-soft state
transition can not happen. For example, the whole accretion flow
remains purely hot (not shown here) throughout the evolution time
for {\bf Run A2} ($\dot M = 0.006\ \dot M_{\rm Edd}$). {\bf Run A2}
is similar to the non-radiative hydrodynamical simulations (e.g.
Stone, Pringle \& Begelman 1999). In other words, the mass accretion
rate is lower than the critical value, thus the radiative cooling of
this model are sufficiently low which results in very weak dynamical
impact.

Fig.\ \ref{fig:mdota1} shows the evolution of the net mass accretion
rate (see Stone, Pringle \& Begelman 1999 for definition) at radius
of $2R_{\rm s}$ (black solid line) and $10R_{\rm s}$ (red dotted
line) in units of Eddington accretion rate for {\bf Run A1}. The net
accretion rate increases by a factor of $\sim 2$ after considering
the radiation cooling. As the gas enters into the central black
hole, the mass accretion rate decreases due to no supplementary
matter. At time  $t \ga $6.5 orbits, the net accretion rate $\sim
0.02\ \dot M_{\rm Edd}$, which is below the critical rate, so the
accretion flow enters into a hard state again. This may be similar
to a soft-to-hard state transition in the decay of BHBs outburst.
Moreover, fluctuation in the mass accretion rate at radius of $R=10\
R_{\rm s}$ is weaker again at time $t \ga $6.5, mainly because of
the disappeared cold clumps.

Fig.\ \ref{fig:A1run} shows the density and temperature of the
accretion system of {\bf Run A1}, from left to right panels
corresponding to time 1.7, 2.0, 4.0 and 8.0 orbits, respectively.
From this plot, obviously there is no conventional cold thin SSD,
but instead of filamentary and sometimes clumpy cold and dense
structures except for the rightmost panel. These cold and dense
structures are concentrated to lower altitude, but clearly they are
highly turbulent, without any sign of settling down onto the
mid-plane. The truncated radius of this model $R_{\rm tr} \approx
35\ R_{\rm s}$ at time $t = 2.0$ (not show picture here) is the same
as {\bf Run A} which has same accretion rate $\dot M \sim 0.03\ \dot
M_{\rm Edd}$ at time t = 8.0.  This result is highly close to the
two-phase LHAF model proposed by Yuan (2003) (see also Xie \& Yuan
2012 and Yang et al. 2015 for more discussions). It is clear that at
time $t=8.0$ the accretion flow becomes hot and dilute again, namely
in hard state.

\subsubsection{{\bf Run A1}: luminous hot accretion flow?}\label{sec:f_factor}

Fig.\ \ref{fig:energya1} shows advection factor of hot accretion
flow at time $t = 1.7$ in {\bf Run A1}. The advection factor is
defined as (Narayan \& Yi 1994; Yuan \& Narayan 2014),
\begin{equation}
f\equiv \frac{Q_{\rm adv}}{Q_{\rm vis}} = 1-\frac{Q_{\rm
rad}^-}{Q_{\rm vis}},
\end{equation}
where $Q_{\rm adv}\equiv \rho \rm d(e/\rho)/\rm
dt+p\nabla\cdot\mathbf{v}\equiv \rho T dS/dt$ ($S$ is the entropy)
is the so-called advection term. In this plot, the curves represent
advection factor $f$ at different angular locations, $\theta =
90\degr$ (black solid; equatorial plane), $\theta = 70\degr$ (blue
dotted) and $\theta = 50\degr$ (red dashed), respectively. The
advection factor is negative at radii larger than $40\ R_{\rm s}$
although the flow is still in hard state. This is exactly the case
of type I luminous hot accretion flow (e.g. Yuan 2001, 2003), which
the radiative cooling is very strong even larger than viscous
heating, but the advection term plays a heating role, rather than a
cooling role, so the accretion flow remains hot.

\subsubsection{radiative cooling in accretion flows}

One advantage of {\bf Run A1} is that, it shows the purely hot
accretion flow (at $t\ga6.5$), the LHAF (at $t\approx 1.7$), and the
two-phase accretion flow (at $1.8 \la t \la 6.5$). We here
investigate in detail the contributions of different radiative
cooling processes. We show in Fig.\ \ref{fig:radA1} the total
radiation cooling rate (multiplied by the volume $R^2 H$, i.e.
$Q^-_{\rm rad} R^2 H$) and the ratio of the Comptonized synchrotron
emission to that of the total radiation ($(Q^-_{\rm syn}+Q^-_{\rm
syn, C})/Q^-_{\rm rad}$) for both the two-phase accretion flow
(right two panels; at $t=1.8-2.0 $) and the ADAF (right two panels;
at $t=7.8-8.0$). The first result we derive from this plot is that,
the Comptonized synchrotron emission dominates for the regions
within $40 R_{\rm s}$, where most of the radiation comes from. This
justifies our argument on the necessity of including synchrotron and
the corresponding Comptonization processes. Besides, the radiation
of hot accretion flow (both ADAF and the hot-phase medium of the
two-phase accretion flow) is spatially extensive, covering a large
volume. We note that, if the magnetic field is relatively stronger
at high latitude regions where the $\beta$ value is smaller compared
to that of the mid-plane (e.g. in MHD simulations by Hawley \&
Krolik 2001; Hirose et al. 2009), then the spatial distribution of
the radiation will be even more extensive and smooth. Additional
emission from the cold clumps make the total emission of the
two-phase accretion flow concentrates towards the equatorial plane.

We should point out that, the concentration of emission site of the
two-phase accretion flow also depends on the wavebands observed. The
hard X-ray emission should still be spatially extensive, as they are
produced by the Compton scattering of the hot electrons. The soft
X-ray and UV emission, on the other hand, will likely concentrates
to the mid-plane, if the clumps are indeed most abundant there.
Currently it remains unclear whether there will be more numerous
smaller clumps formed, and whether these clumps can be supported to
stay at higher altitude by the magnetic field. Further MHD
simulations with higher spatial resolution will be required to
clarify these questions.

\begin{figure*}
\begin{center}
\includegraphics[width=0.85\textwidth]{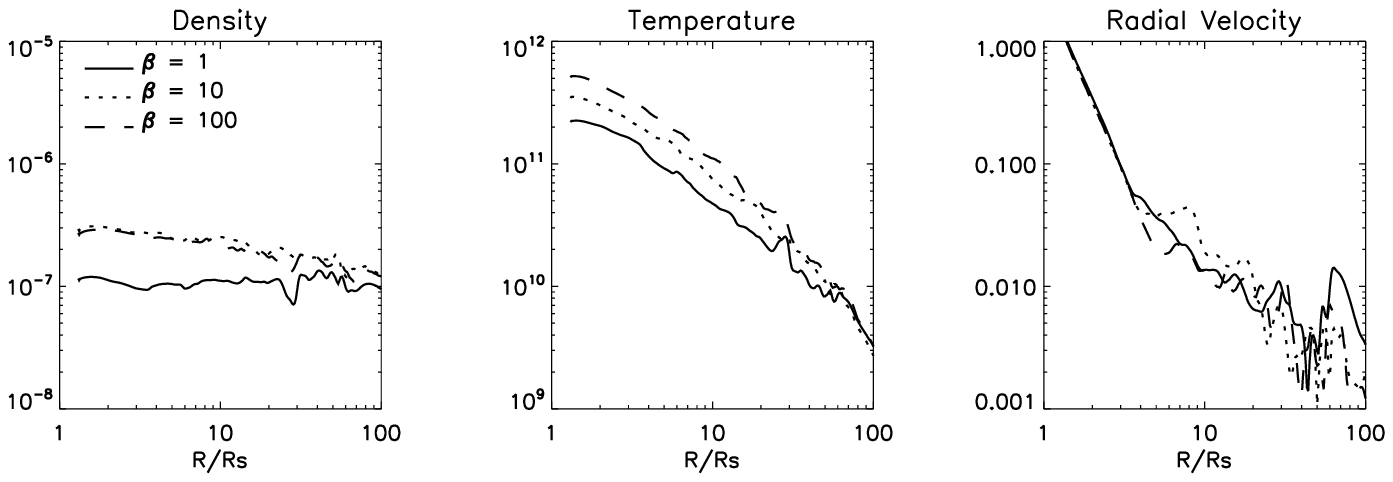}\\
\includegraphics[width=0.85\textwidth]{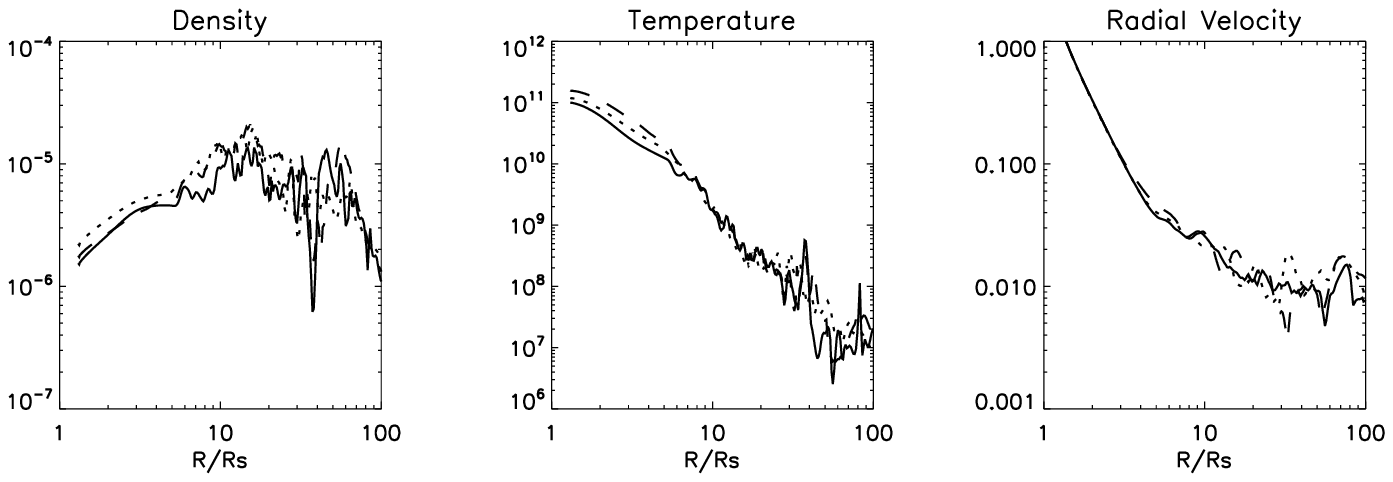}
\caption{Comparison of the properties of simulations with different
$\beta$ values, i.e. the solid, dotted and dashed curves correspond
to models with $\beta = 1, \beta = 10$ and $\beta = 100$,
corresponding to {\bf Run B1}, {\bf Run A} and {\bf Run B},
respectively. The upper panels show the averaged quantities
(density, \textbf{temperature} and radial velocity) at the mid-plane
over the period $1.3 - 1.4$ (without radiative cooling), while
bottom panels show averaged quantities over the period $1.8 - 2.0$
(with radiative cooling).} \label{fig:beta}
\end{center}
\end{figure*}

\subsection{Effects of plasma parameter $\beta$}\label{sec:beta}

In order to investigate the effect of magnetic field strength, we
also run two additional models with $\beta=100$ ({\bf Run B}) and
$\beta=1$ ({\bf Run B1}). The other model parameters are the same as
those of the fiducial run. Fig.\ \ref{fig:beta} shows the profiles
of density, ion temperature and radial velocity at the mid-plane,
before (averaged over the period $t =1.3 - 1.4 $; top panels) and
after (averaged over the period $t= 1.8 - 2.0 $; bottom panels)
including the radiative cooling.

We first analyze the results without radiative cooling. If magnetic
field is weak ({\bf Run A} and {\bf Run B}), the accretion flow will
be gas pressure dominated. Consequently, there will be no
significant differences between these simulations. The ion
temperature of {\bf Run B} is slightly higher, mainly because the
gas pressure (roughly $\propto \rho T_{\rm i}$) increases as $\beta$
increases, as the total pressure remains less affected. If magnetic
field becomes strong enough, i.e. $\beta\sim 1$, its dynamical
impact will be evident, even without considering the radiation
processes. Indeed, When comparing {\bf Run B1} with {\bf Run A}, we
find that the density becomes lower in low $\beta$ case. This is
because, we additionally put the magnetic pressure during the
initial setup of the equilibrium (gravity against gas pressure
gradient force) torus. Consequently, more fraction of the accreting
material will be driven out as outflow due to the enhanced pressure
gradient force at the early stage of evolution. The density will be
lower and the accretion rate will be reduced. Besides, the ion
temperature of {\bf Run B1}, whose accretion rate is lower, is lower
compared to that of {\bf Run A}. This is because of weaker
compression work ($p{\rm d} (1/\rho)$, which is proportional to
density gradient; cf. the density profile in Fig.\ \ref{fig:beta})
done onto the gas in {\bf Run B1}.
\begin{figure*}
\begin{center}
\includegraphics[width=0.40\textwidth]{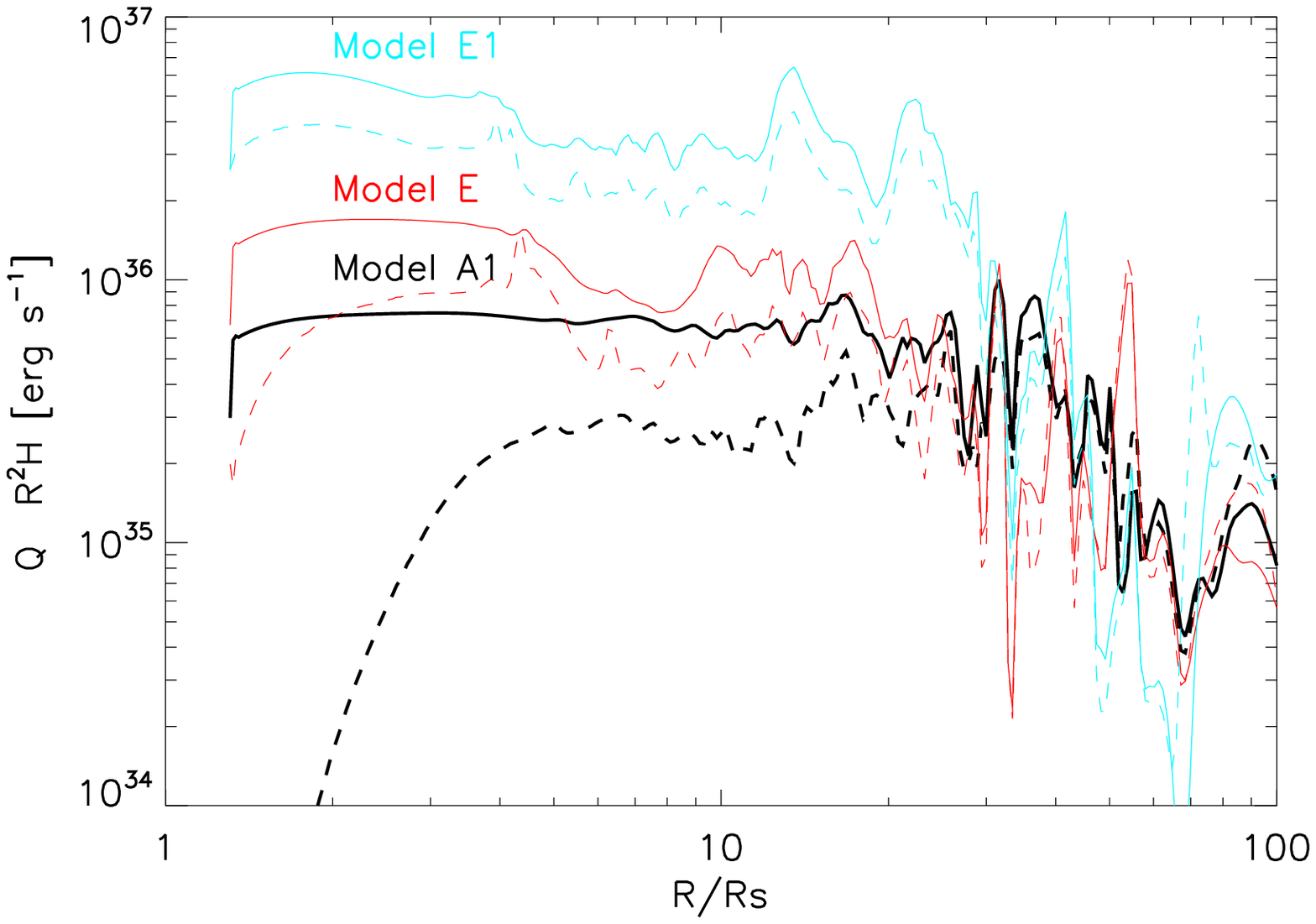}
\includegraphics[width=0.55\textwidth]{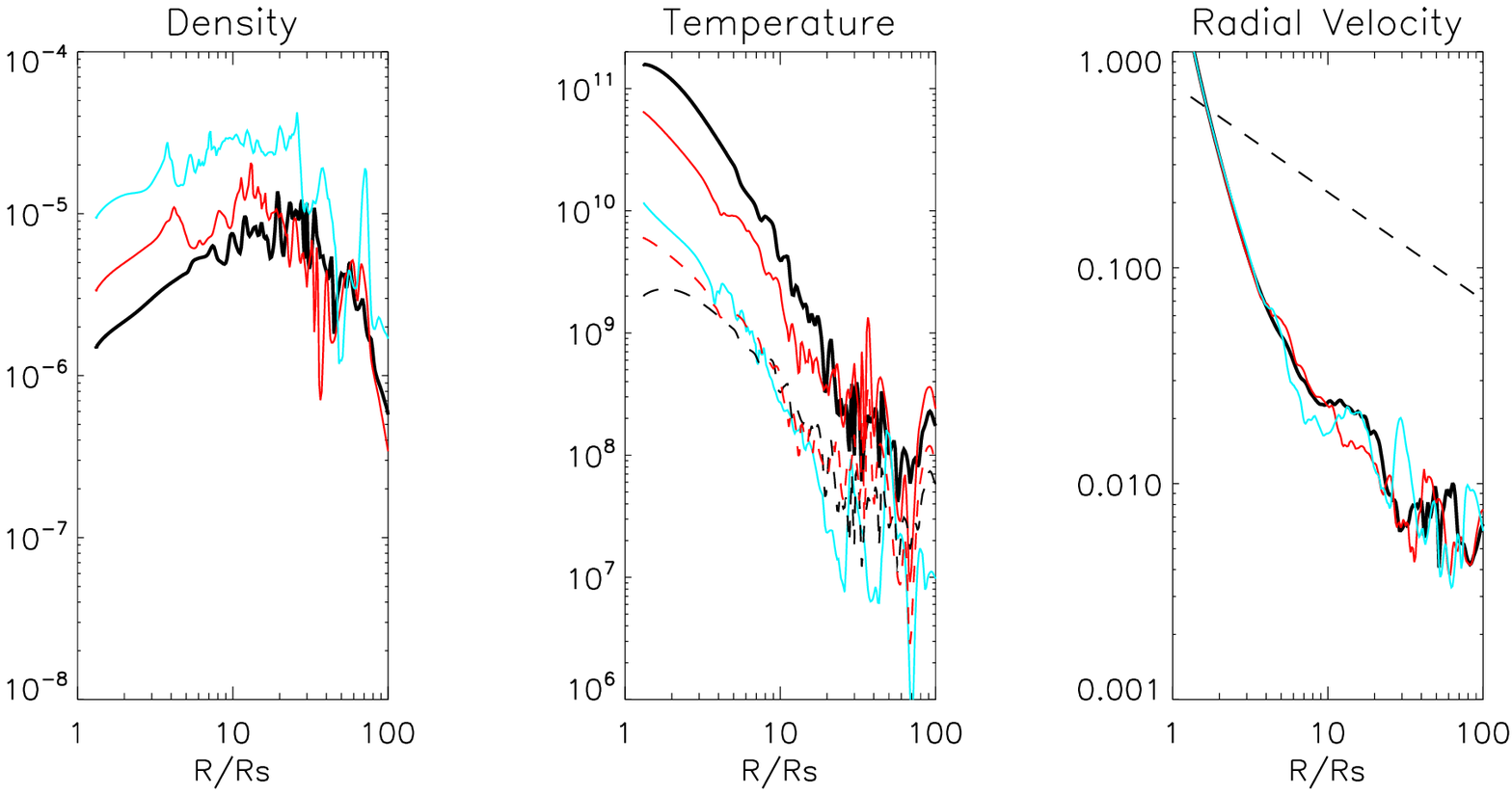}
\caption{Comparison of simulations with different $T_{\rm e}/T_{\rm
i}$ values over period of $t$= 1.8 - 2.0 orbits. As labeled in the
figure, {\bf Run A1}, {\bf Run E} and {\bf Run E1} have $k=1$,
$k=0.5$ and $T_e = T_i$ (sole temperature simulation), while they
share the same other initial conditions. In the leftmost panel, the
solid and dashed curves are the viscous heating rate and the total
radiative cooling rate (both are multiplied by $R^2\ H$). In the
right three panels, show the density, temperature (solid line
denotes the ions temperature, the dashed lines denotes the electrons
temperature) and radial velocity in units of light speed (the dashed
line denotes free fall speed.)}\label{fig:tie}
\end{center}
\end{figure*}

We then compare results with and without radiative cooling. The
synchrotron emission, which provides the seed photons for the
Compton scattering process, is more sensitive to the electron
temperature than the magnetic field strength, i.e. we roughly have
$Q_{\rm syn}\propto T_{\rm e}^7 B^3\propto T_{\rm e}^7 P_{\rm
mag}^{3/2}$ (This can be derived from Eq. 8, and note that $\nu_c$
is roughly proportional to $B T_{\rm e}^2$; NY95; See also Mehadevan
1997 for the dependence on $T_{\rm e}$). In this sense, The impact
of including radiative cooling in {\bf Run B}, whose electron
temperature is the highest before considering radiative cooling (cf.
top middle panels of Fig.\ \ref{fig:beta}), is significant, while
the impact of including radiative cooling in {\bf Run B1} is much
weaker, especially on the temperature of the gas. The difference of
the total radiative cooling rates of the three runs are highly
reduced, i.e. they differ by factors of $\la 2$ for our chosen
parameters. We note that the similarities of the three runs after
including radiative cooling may be a coincidence, and should not be
taken seriously. Radiative MHD simulations are required for a better
understanding on the impacts of the magnetic fields, both
dynamically (through magneto-rotational instability) and radiatively
(through synchrotron emission).

Moreover, we also point out that our approach can only mimic the
randomly-oriented turbulent magnetic fields, but not the global
large scale one. The non-radiative simulations are different from
those non-radiative MHD simulations, e.g. Stone \& Pringle (2001).
Their results show that the mass accretion rate depends on the
initial magnetic field strength, i.e. stronger magnetic field will
lead to higher mass accretion rate. The reason is that, the Maxwell
stress, which determines the capability of transporting angular
momentum, will be stronger for cases with stronger vertical fields.
In other words, stronger magnetic fields will generally have larger
`effective' $\alpha$, while it is fixed to a constant in our HD
simulations. Furthermore, in the MHD case, when radiative cooling
becomes dominant, ADAF-like gaseous disk is shrieked in the vertical
direction with frozen-in magnetic field, and magnetic pressure
becomes relatively dominant in the cool disk. Hirose et al. (2009)
showed that the plasma parameter $\beta$ in coronal region became
lower value than the disk region when radiative cooling becomes
important. Therefore, the assumption of constant $\beta$ will
under-estimate synchrotron radiation.

\subsection{The dependence on temperature ratio $T_{\rm
e}/T_{\rm i}$}

Obviously, only when radiative cooling becomes important in
determining the dynamical structure of the accretion flow, the
simulation results will be sensitive to the value of $T_{\rm
e}/T_{\rm i}$ (Mo{\'s}cibrodzka et al. 2009; Dibi et al. 2012).

From theoretical point of view, $T_{\rm e}/T_{\rm i} $ depends
mainly on two quantities. The first is the accretion rate. At
sufficiently low accretion rate (e.g. the quiescent state and also
the low luminosity hard state), the
radiative cooling of electrons is insignificant, thus the $T_{\rm
e}/T_{\rm i}$ will be moderate (it depends on the energy
deposition and also the adiabatic index differences between
electrons and ions; cf. Yuan \& Narayan 2014). As the accretion
rate increases (i.e. normal hard state), the electrons will suffer
additional radiative cooling
and $T_{\rm e}$ will be reduced. $T_{\rm i}$, on the other hand,
remains nearly unaffected, mainly because the Couloumb
collision is small compared to the viscous heating to ions at such
moderate accretion rates. In other words, the value of $T_{\rm
e}/T_{\rm i}$ will become smaller in this regime. When the accretion
rate is very high (e.g. the bright hard state and the intermediate
state, i.e. the LHAF regime), the Couloumb coupling between
electrons and ions, which scales as $\rho^2\ (T_{\rm i} - T_{\rm
e})$, is so strong that the temperature differences between
electrons and ions are reduced, i.e. $T_{\rm e}/T_{\rm i}$ will
become moderately large again (see e.g. Yuan 2001; Xie et al.
2010). Note that for cases of SSD and possibly the cold clumps of
the two-phase accretion flow, we will have $T_{\rm e} = T_{\rm i}$.
Observationally we do observe an
anti-correlation between electron temperature and the bolometric
luminosity for the outburst of black hole X-ray binaries (e.g.
Joinet et al. 2008; Miyakawa et al. 2008; Motta et al. 2009;
Natalucci et al. 2014). The second quantity is the fraction of
viscous heating that goes to electrons directly (see e.g. Xie \&
Yuan 2012 for a brief summary on constrains of this value). For
given density and accretion rate, if more fraction of viscous
heating goes to the electrons, obviously it will have a higher value
of $T_{\rm e}/T_{\rm i}$, and the corresponding bolometric
luminosity will also be enhanced.

We run two new simulations to investigate the impact of temperature
ratio. With other parameters the same to those of {\bf Run A1}, we
set in {\bf Run E} $k=0.5$ and in {\bf Run E1} $T_{\rm e} = T_{\rm
i}$. {The densities are high in these runs, and the radiative
cooling is dynamically important, representing the bright hard (and
also intermediate) state. As radiative cooling is triggered,
differences are evident, as shown in Fig.\ \ref{fig:tie} when a
quasi-steady state is reached.

Reducing the differences between $T_{\rm e}$ and $T_{\rm i}$ (as in
{\bf Run E1} and {\bf Run E}, compared to {\bf Run A1}) is
equivalent to enhance the coupling between electrons and ions. The
$T_{\rm e}$ will then be increased, while the $T_{\rm i}$ will be
reduced. The scale height of the hot accretion flow ($\propto
(T_{\rm e}+T_{\rm i})^{1/2}$) will also be reduced, which leads to
an enhancement in gas density. The radiative cooling rate
(synchrotron and its Comptonization, bremsstrahlung) will also be
increased. All these results, as shown in Fig.\ \ref{fig:tie}, are
easy to understand.

In the case of {\bf Run E} series they start to form clumpy
structures at lower accretion rate, i.e. they have more lower
critical accretion rate of the purely hot accretion flow compared to
those with $k=1$ models. We also run some other simulations (not
shown here) and the results confirm this point, such as the critical
accretion rate is reduced to  $\sim 0.006\ \dot M_{\rm Edd}$ in {\bf
Run E1}. The ratio between electron and ion temperature is the key
point for the radiative cooling process, but it is very complex and
depends on many factors and microphysical processes. Two-fluid
(electrons and ions) radiative (MHD) simulations are needed to solve
this problem. Only under this approach, we can investigate impact of
the fraction of viscous heating to electrons, and come out a more
reasonable $T_{\rm e}/T_{\rm i}$ value (i.e. varies at different
location and/or time).

\section{Summary and Discussions}\label{summary}
In this paper, we present a number of two-dimensional hydrodynamical
simulations of accretion flow on to a black hole, focusing on the
impact of radiation cooling and state transitions. Compared to
previous works (e.g. Das \& Sharma 2013), our technical improvements
includes: (1), the hot accretion flow in our simulations is
two-temperature (NY95); (2), we follow the treatment of NY95 to
consider the synchrotron emission and the corresponding Compton
up-scattering process (see also Fragile \& Meier 2009), which are
crucial to generate the radiation observed in black hole X-ray
binaries.

We summarize our major results as follows. Firstly, as the accretion
rate increases, the truncated radius moves inward. This result is
consistent with theoretical expectation (Esin et al. 1997) and the
numerical simulations by Das \& Sharma (2013). Secondly, when the
accretion rate reaches the critical value $\dot M = 0.03\ \dot
M_{\rm Edd}\approx3\alpha\ \dot M_{\rm Edd}$ (Xie \& Yuan 2012), we
observe the formation of clumpy/filamentary structures within the
hot medium, i.e. the whole accretion flow is likely to be in a
two-phase accretion mode (see, e.g. Yuan 2003; Xie \& Yuan 2012). As
the accretion rate increases, these clumps grow and/or merge.
Eventually the turbulent motions of the hot gas can not support them
any longer, and they will settle down onto the mid-plane and form a
thin disc. In other words, the clumpy accretion mode do exist before
the eventually formation of the disc-corona configuration. It is
speculated that this accretion mode may be responsible for the
intermediate state of X-ray binaries (e.g. Yang et al. 2015).

We still lack direct evidences for the existence of cold clumps (or
alternatively filaments, clouds), especially in BHBs. In AGNs, some
clues can be derived from variability studies. For example, in a
broad absorption line (BAL; outflowing gas with velocity $\sim 0.1\
c$) quasar monitoring campaign, Capellupo et al. (2012) found that
the variability favours the idea that those absorptions varies (on
timescale of years) because of the movements of individual
outflowing clouds (or at least substructures in the moving flow).
Another example comes from NGC 5548 (Kaastra et al. 2014), where
clumpy streams of ionized hot gas blocking the emission from the
nucleus are observed.

Below we will highlight several notable shortages of our work, and
discuss their consequences.

\subsection{Oversimplification in the radiative cooling}
In this work, we highly simplify the calculation of
radiative cooling. The first is that, we use the critical frequency
$\nu_c$ of the self-absorbed synchrotron emission to estimate the
synchrotron cooling rate. Besides, we also use this frequency as the
representative seed photon frequency for the Compton scattering
process. Such approximation, which avoids any integration (in
frequency $\nu$), highly accelerates the computational speed.
However, its accuracy is difficult to constrain.

More importantly, our Comptonization is based on the local Compton
scattering approximation. In the optically thin medium as the case
of hot accretion flow, seed photons can propagate a long distance
from one location to another, and scatter with the hot electrons
there. Such ``global'' Compton scattering process is not taken into
account in our simulations. The global Comptonization process is
recently examined in one-dimensional calculations, and it will make
the electron temperature profile more flattened and the absolute
$T_{\rm e}$ value reduced (by a factor of $\sim 1/3$. see e.g. Yuan
, Xie \& Ostriker 2009; Xie et al. 2010; Nied{\'z}wiecki et al.
2012). If we do not want to employ the computationally-expensive
Monte Carlo technique, then one easy way to consider this effect is
to create a correction factor profile from these numerical
calculations and then apply it to the simulations here. However,
such treatment is also very crude, and we omit it in this work.

\subsection{Magnetic fields (large-scale ordered) in the hot accretion flow}\label{sec:mag}

In this work we used plasma $\beta$ parameter to consider the
effects of magnetic field (and correspondingly the synchrotron
emission). Such approach has several limitations, which we will
discuss below.

Firstly, this approach can not properly take into account the
large-scale ordered magnetic fields. Plasma $\beta$ can only provide
a feasible and reasonable approximation for the consideration of the
tangled, turbulent magnetic fields, but even in this case, it is not
a constant within the accretion flow (Hawley \& Krolik 2001).
Besides, the dynamical structure of hot accretion flow will be
highly revised if such large-scale magnetic field exist (Machida et
al. 2006; Fragile \& Meier 2009).  The ``effective'' $\alpha$ will
be highly enhanced if the accretion flow has large net magnetic flux
(cf. Penna et al. 2013; Bai \& Stone 2013). Besides, the large scale
magnetic field may also be crucial for the formation and
acceleration of the Poynting-dominated jet powered by the black hole
spin (Blandford \& Znajek 1977; Tchekhovskoy, Narayan \&  McKinney
2011), the disk-jet powered by the disk rotation (Yuan \& Narayan
2014; Yuan et al. 2015), or jet powered by the radiation from
supercritical accretion flow (Sadowski \& Narayan 2015).

Secondly, one evolutionary stage of our simulations (e.g. in {\bf
Run A1}), before the eventual collapse into a cold thin disc, is the
formation of the cold and dense clumpy/filamentary components. If
these cold clumps are still coupled to the magnetic fields (valid
likely only for small clumps), then the magnetic fields may service
as natural barrier for the merger of those small clumps, i.e. the
clumps will be confined by the large-scale magnetic fields. Besides,
the large scale magnetic fields will provide additional vertical
support for those cooling gas components, thus maintain them at
higher altitude (Machida et al. 2006).

Finally, one interesting property related to the accumulation of the
large scale global magnetic fields is that, they may help on the
understanding of the the hysteretic cycle of black hole state
transitions (Begelman \& Armitage 2014 and references therein for
alternative scenarios), i.e. the critical luminosity of the
hard-to-soft transition is generally brighter than that of the
soft-to-hard state. In the scenario proposed by Begelman \& Armitage
(2014), they expect that, during the hard state, vertical magnetic
flux, which is generated near the truncated radius $R_{\rm tr}$, is
advected inward and accumulates stochastically within the inner hot
flow. Consequently the $\alpha$ will be large, leading to a higher
critical luminosity for the hard-to-soft state transition (see also
Yu et al. 2015 for the stability analysis of low-$\beta$ hot
accretion flows). On the other hand, the magnetic flux is gradually
diffused out in the cold SSD during the soft state, thus the
critical luminosity of the soft-to-hard state transition will be
much lower. The numerical realization of this scenario is beyond the
scope of current work.

\subsection{Turbulent dissipation and corona heating}

Another issue is related to the turbulent/viscous dissipation within
the accretion flow. In this work (and also most of the
hydrodynamical simulations of accretion flows), we adopt the
$\alpha$-viscosity prescription (Shakura \& Sunyaev 1973), which
assumes that the dissipation heating rate is proportional to the
density of the gas. One consequence of this assumption is that, it
assumes most of the ``viscous'' heating happens close to the
mid-plane of the accretion flow.

Contradict to this assumption, recent MHD simulations, in which the
magnetic reconnection is considered as the ``physical'' heating
mechanism of the accretion disc, indicate that the vertical
distribution of energy dissipation actually do not follow the
density profile, but can extend to much higher height (e.g. Jiang,
Stone \& Davis 2014 and references therein). If this is indeed the
case, then it will provide a natural heating mechanism for the
corona in those disc-corona models (see e.g. Liu et al. 2002; Cao
2009), where they generally assume a significant fraction (of order
$\sim 50\%$) of the ``viscous'' heating energy goes into the corona
region. Besides, if such differences are taken into account, the
mid-plane regions will be cooler, while the high altitude regions
should be hotter.

\section{ACKNOWLEDGMENTS}
M.C.W. thanks Dr. De-Fu Bu,  Dr. Zhen-Yi Cai and Prof. Feng Yuan for
useful discussions. This work was supported in part by the Natural
Science Foundation of China (grants U1231106, U1431228, 11133005,
11203057, 11233003, 11273042 and 11421303), the National Basic
Research Program of China (2012CB821801), the Strategic Priority
Research Program of the Chinese Academy of Sciences (XDB09000000),
and the grant from `the Fundamental Research Funds for the Central
Universities'. This work made use of the High Performance Computing
Resource in the Core Facility for Advanced Research Computing at
Shanghai Astronomical Observatory. F.G.X. was also supported in part
by the Youth Innovation Promotion Association of CAS (id. 2016243).

\end{document}